\documentclass[a4paper,11pt]{article}
\usepackage{pos}

\title{Recent Progress in Lattice Parton-Distribution Calculations}

\ShortTitle{Recent Progress in Lattice Parton-Distribution Calculations}

\author*[a,b,1]{Huey-Wen Lin}
\affiliation[a]{Department of Physics and Astronomy, Michigan State University,\\
        East Lansing, MI, 48824, U.S.A}

\affiliation[b]{Department of Computational Mathematics,
        Science and Engineering, Michigan State University,\\
        East Lansing, MI, 48824, U.S.A}
\emailAdd{hwlin@pa.msu.edu}
\note{The work of HL is partially supported by the US National Science Foundation under grant PHY 1653405 “CAREER: Constraining Parton Distribution Functions for New-Physics Searches” and by the Research Corporation for Science Advancement through the Cottrell Scholar Award ``Unveiling the Three-Dimensional Structure of Nucleons''.}

\abstract{
The large‐momentum effective theory (LaMET) framework has been widely used to determine the Bjorken‐$x$ dependence of parton distribution functions (PDFs) in lattice‐QCD hadron-structure calculations.
In this talk, I highlighted selected recent lattice-QCD results on parton distributions from MSULat group.
We use clover valence fermions on ensembles generated by MILC Collaboration with lattice spacing $a\approx 0.06$, 0.09, and 0.12~fm, with $M_\pi L \approx 4$, and with pion masses including 135, 220 and 310~MeV and $N_f=2+1+1$ flavors of highly improved staggered dynamical quarks.
Results include the continuum-physical isovector nucleon PDF, a first study of the strange and charm PDFs and the pion and kaon valence-quark PDFs.
We also reported results on the $Q^2$ and $x$ dependence of nucleon isovector unpolarized and helicity GPDs calculated directly at physical pion mass;
we also make a comparison of the GPDs with the traditional moment methods from other lattice calculations.
}

\FullConference{%
 The 38th International Symposium on Lattice Field Theory, LATTICE2021
  26th-30th July, 2021
  Zoom/Gather@Massachusetts Institute of Technology
}

\begin{document}
\maketitle

\section{Introduction}
\vspace{-0.3cm}
There has been exciting progress in the past decade on an increasing number of lattice hadronic structure calculations at the physical pion mass, with many calculations now coming with high statistics ($O(100\text{k})$ measurements) and some with multiple lattice spacings and volumes to control lattice artifacts.
Furthermore, breakthroughs have been made in $x$-dependent methods, such as large-momentum effective theory (LaMET)~\cite{Ji:2013dva,Ji:2014gla,Ji:2020ect} (quasi-PDFs),
with pioneering works showing great promise in obtaining quantitative results for the unpolarized, helicity and transversity quark and antiquark distributions~\cite{Lin:2014gaa,Lin:2014yra,Lin:2014zya,Chen:2016utp,Alexandrou:2014pna,Alexandrou:2015rja}
using the quasi-PDFs approach~\cite{Ji:2013dva}.
Increasingly many lattice works are being performed at physical pion mass since the first study in Ref.~\cite{Lin:2017ani}.
A recent review of the theory and lattice calculations can be found in Ref.~\cite{Ji:2020ect}.

In this proceeding, we focus on work done by our group (MSULat) since the last lattice conference.
In these calculations, we use clover valence fermions on ensembles generated by MILC Collaboration~\cite{MILC:2010pul,Bazavov:2012xda} with lattice spacing $a\approx 0.06$, 0.09, and 0.12~fm, with $M_\pi L \approx 4$, and with the physical pion mass ranging from $135$, 220 and 310~MeV and $N_f=2+1+1$ (degenerate up/down, strange and charm) flavors of highly improved staggered dynamical quarks (HISQ)~\cite{Follana:2006rc}.
The gauge links are one-step hypercubic (HYP) smeared~\cite{Hasenfratz:2001hp} to suppress discretization effects.
The clover parameters are tuned to recover the lowest sea pion mass of the HISQ quarks.
The ``mixed-action'' approach is commonly used, and there is promising agreement between the lattice-calculated nucleon charges, moments and form factors and the experimental data when applicable. (See Ref.~\cite{Mondal:2020cmt} and references within.)
Gaussian momentum smearing~\cite{Bali:2016lva} is used on the quark field to improve the overlap with ground-state nucleons of the desired boost momentum, allowing us to reach higher boost momentum for the nucleon states.
At each boost momentum, the nucleon energy is obtained through a two-state fit to the two-point correlator,
$C_\text{2pt}(t)=|\mathcal{A}_0|^2e^{-E_0 t}+|\mathcal{A}_1|^2e^{-E_1 t}+\dots$,
where $E_i$ and $A_i$ are the energy and overlap factor between the lattice nucleon operator and desired state $|i\rangle$,
and $i=0$ ($i=i$) stands for the ground (excited) state.
The three-point correlator needed to extract the ground-state matrix elements of nucleon and mesons, $\langle 0|O|0 \rangle$, is then
\begin{align}
    C_{\text{3pt}}(t_\text{sep},t)=&|A_0|^2\langle0|O|0\rangle e^{-E_0 t}e^{-E_0 (t_\text{sep}-t)}+
    |A_0||A_1|\langle1|O|0\rangle e^{-E_0 t}e^{-E_1 (t_\text{sep}-t)}+\nonumber\\
    &|A_1||A_0|\langle0|O|1\rangle e^{-E_1 t}e^{-E_0 (t_\text{sep}-t)}+
    |A_1|^2\langle1|O|1\rangle e^{-E_1 t}e^{-E_1 (t_\text{sep}-t)}+\dots .
\end{align}

\section{Nucleon PDFs}
\begin{figure}[htbp]
\includegraphics[width=.32\textwidth]{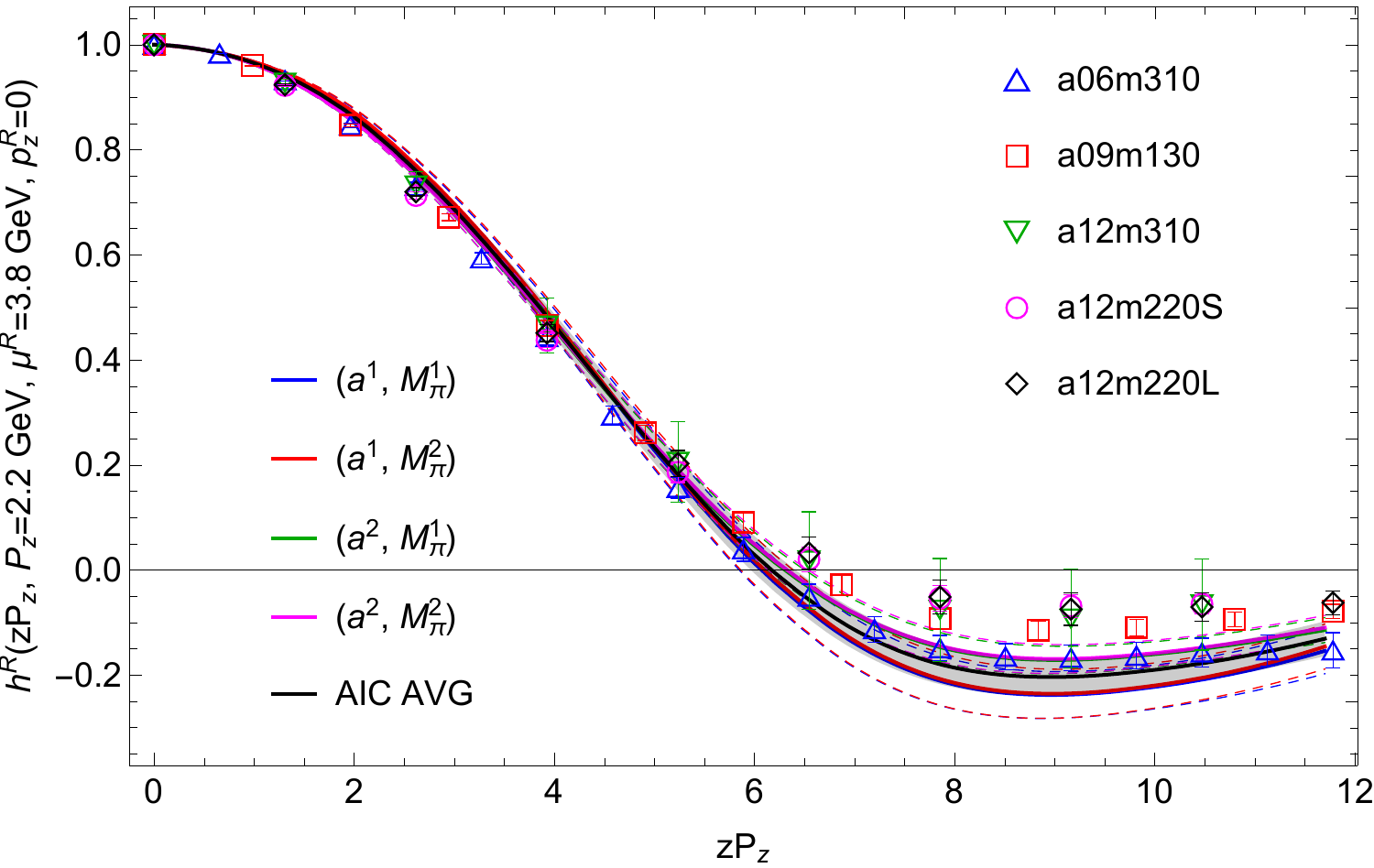}
\includegraphics[width=.32\textwidth]{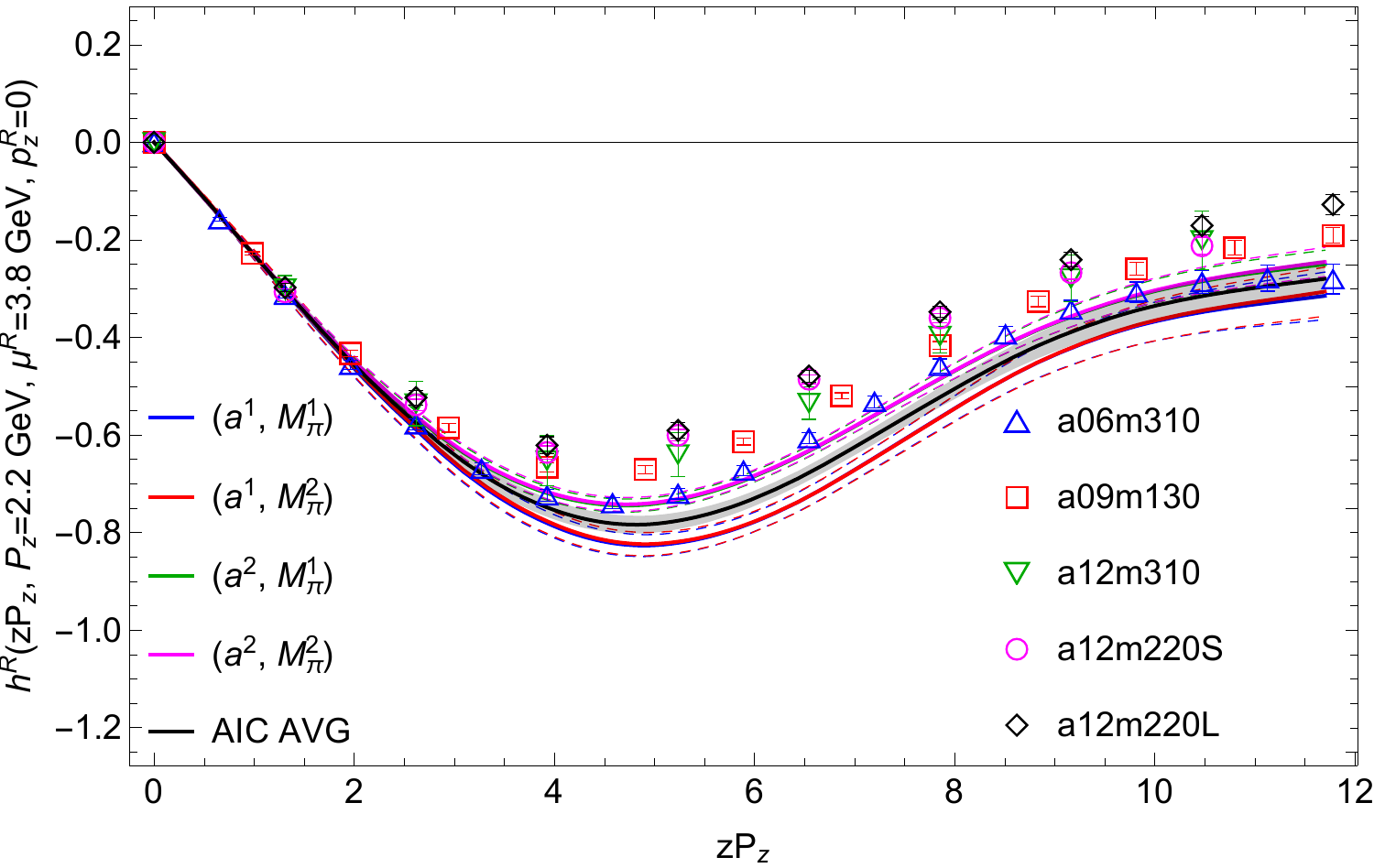}
\includegraphics[width=.32\textwidth]{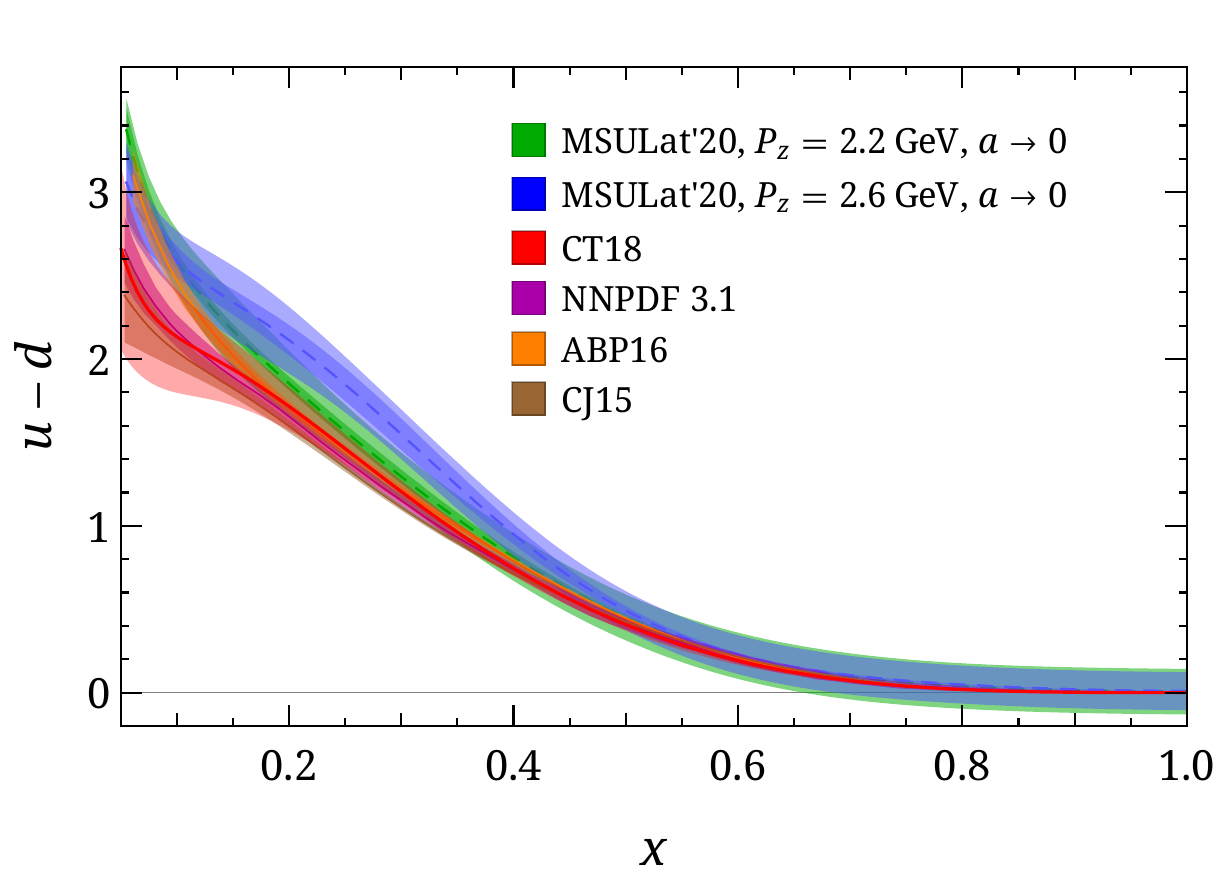}
\caption{
The physical-continuum extrapolation of the real (left) and imaginary (middle) matrix elements from the ensembles with nucleon boost momentum around 2.2~GeV~\cite{Lin:2020fsj}.
Various ans\"{a}tze with linear/quadratic extrapolation in lattice spacing and pion mass are shown as solid lines for the central values and dashed lines for uncertainties.
The filled band shows the AIC-averaged physical-continuum matrix elements.
(right)
The nucleon isovector unpolarized PDFs from our lattice calculation in the physical-continuum limit compared with global fits from Refs.~\cite{Hou:2019efy,Alekhin:2017kpj,Ball:2017nwa,Accardi:2016qay}.
}
\label{fig:P2GeV}
\end{figure}

\vspace{-0.3cm}
The most studied $x$-dependent structures are the nucleon unpolarized isovector parton distribution functions (PDFs).
In Ref.~\cite{Lin:2020fsj}, we present the first lattice-QCD calculation of the nucleon isovector unpolarized PDFs in the physical-continuum limit.
The lattice results are calculated using ensembles with multiple sea pion masses with the lightest one around 135~MeV, three lattice spacings $a\in[0.06,0.12]$~fm, and multiple volumes with $M_\pi L$ ranging 3.3 to 5.5.
We perform a simultaneous chiral-continuum extrapolation to obtain RI/MOM renormalized nucleon matrix elements with various Wilson-link displacements
and obtained four physical-continuum matrix elements (linear/quadratic in lattice spacing and pion mass), as shown in solid (central value) and dashed (error band) lines in Fig.~\ref{fig:P2GeV}.
We find that the four fits from the real matrix elements are in good agreement, whereas more fluctuations are seen in the imaginary matrix elements.
The fluctuations are mainly dominated by the lattice-spacing extrapolation.
Using a linear lattice-spacing extrapolation form results in slightly higher continuum-limit matrix elements than those obtained from a quadratic form.

We then apply one-loop perturbative matching to the quasi-PDFs to obtain the lightcone PDFs, and compare our results with a selection of global-fit PDFs.
Note that lattice discretization systematics were not taken into account up to this point, and the twist-4 effects were assumed to be $\mathcal{O}(\Lambda_\text{QCD}^2/P_z^2)$.
The twist-4 systematic is negligible, since few-percent effects at this large momentum are much smaller than the statistical and other systematic uncertainties.
Since we account for these neglected systematics in this work, the total uncertainty appears larger than those of previous quasi-PDF works, even though the statistical error remains comparable.
When comparing our continuum-physical nucleon isovector PDFs with those obtained from global fits,
CT18NNLO~\cite{Hou:2019efy},
NNPDF3.1NNLO~\cite{Ball:2017nwa},
ABP16~\cite{Alekhin:2017kpj}, and
CJ15~\cite{Accardi:2016qay},
we found our results, even with only the errors considered by inner statistical bands, have nice agreement.
The errors increase toward the smaller-$x$ region for both lattice and global fitted PDFs, but overall, they agree within two standard deviations.

\begin{figure}[tb!]
\centering	
\includegraphics[width=0.45\linewidth]{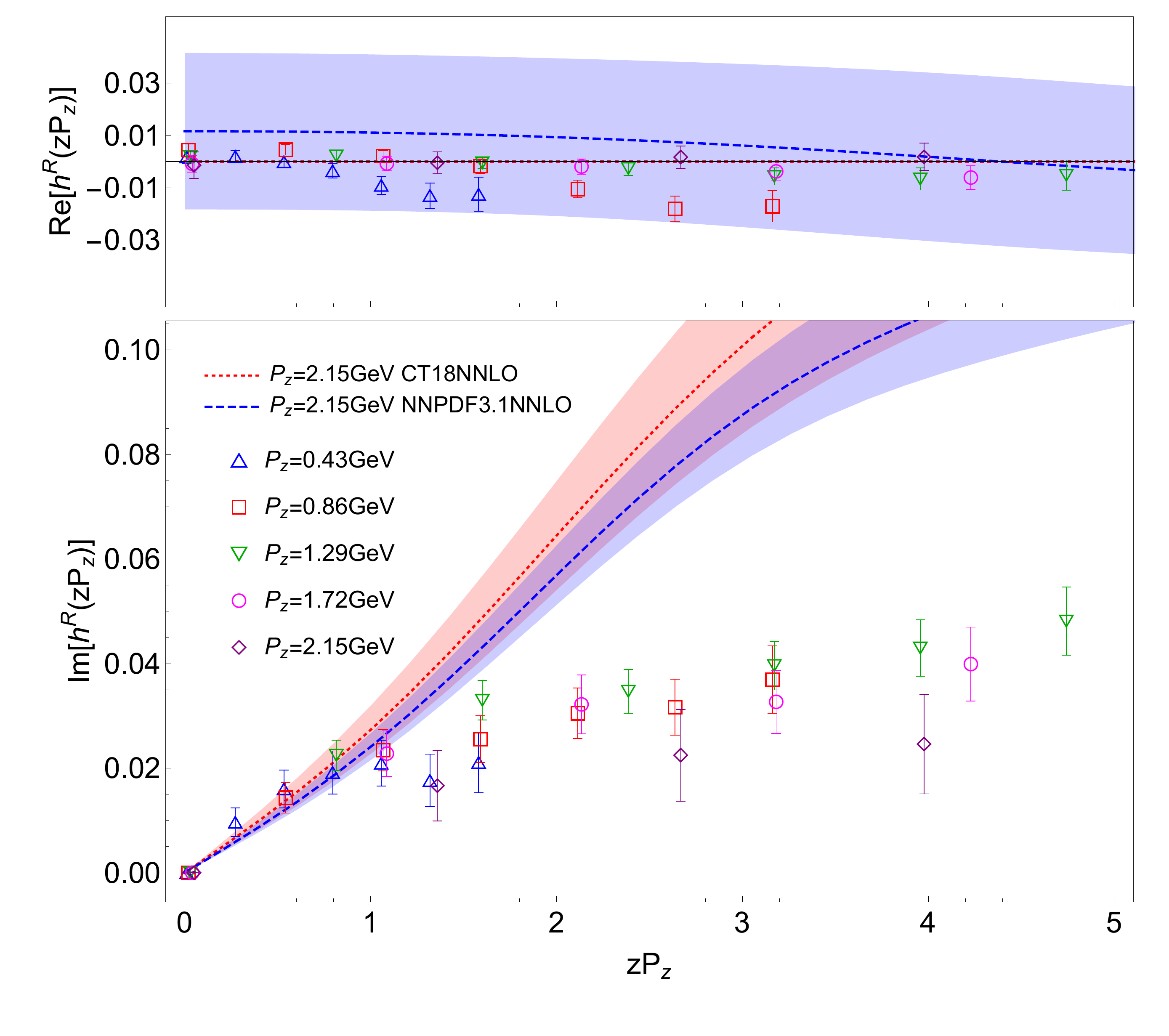}
\includegraphics[width=0.45\linewidth]{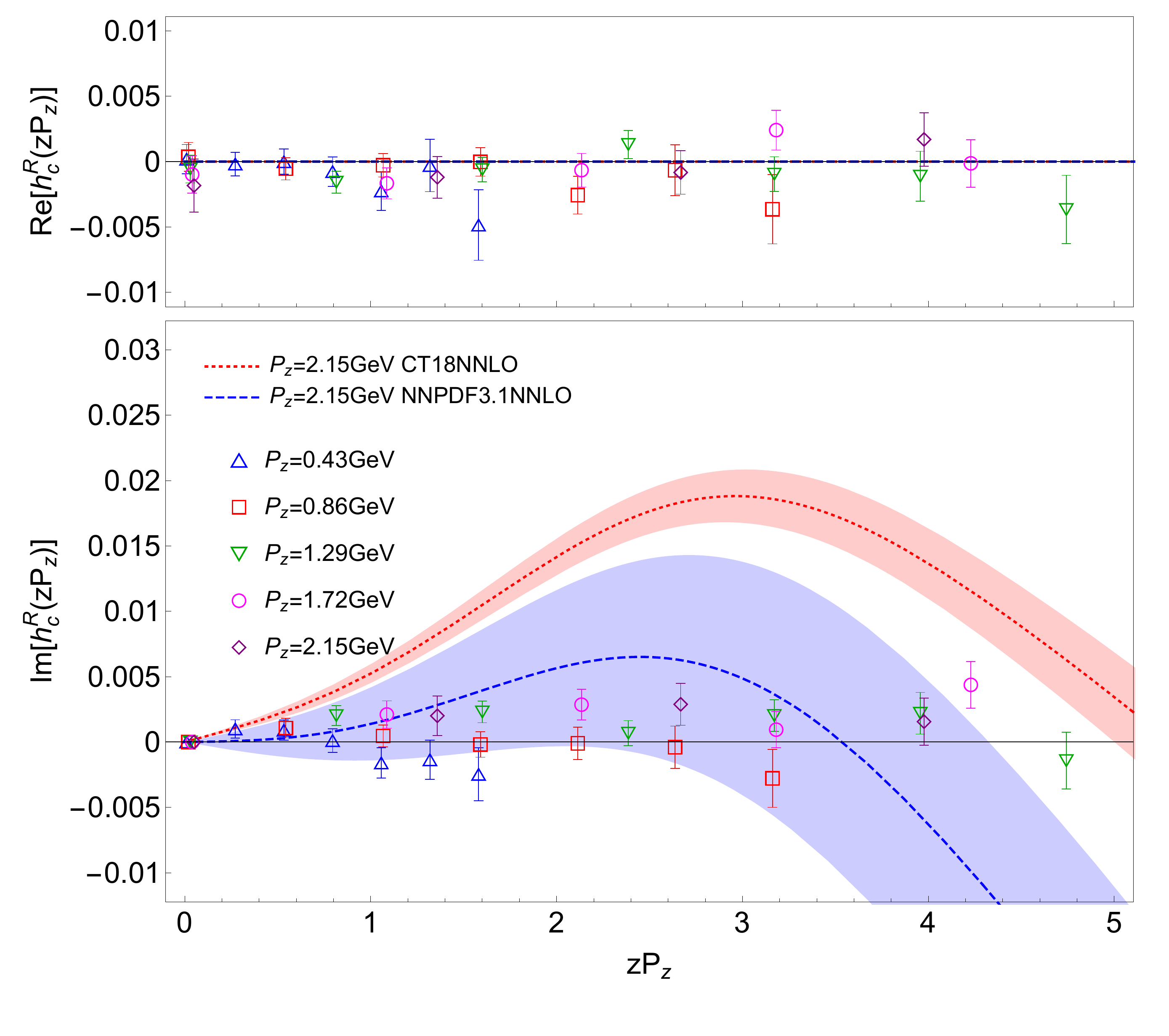}
\caption{\label{fig:strange-charm-PDFs}
The real (top) and  imaginary (bottom) parts of the strange (left) and charm (right) quasi-PDF matrix elements in coordinate space from our calculations at physical pion mass with $P_z \in [0.43, 2.15]$~GeV~\cite{Zhang:2020dkn}, along with those from CT18 and NNPDF NNLO in RI/MOM renormalized scale of 2.3~GeV.
For the strange results:
The CT18 analysis assumes $s(x)=\bar{s}(x)$, so their results are exactly zero after matching and Fourier transformation.
Our real matrix elements at $P_z>1$~GeV are consistent with zero, supporting strange-antistrange symmetry, while our imaginary ones are smaller than global-fit results.
For the charm results:
The real part is consistent with zero, while the imaginary part is within the bounds of NNPDF3.1, but smaller than CT18 results.
}
\end{figure}

In Ref.~\cite{Zhang:2020dkn}, we report the first lattice-QCD calculations of the strange and charm parton distributions using LaMET approach were performed using a 310-MeV HISQ ensemble with single lattice spacing 0.12~fm.
The calculation of light ($M_\pi\approx310$~MeV) and strange nucleon ($M_\pi\approx690$~MeV) two-point correlators includes 344,064 (57,344)
measurements in total.
We found that our renormalized real matrix elements are zero within our statistical errors for both strange and charm, supporting the strange-antistrange and charm-anticharm symmetry assumptions commonly adopted by most global PDF analyses, as shown in Fig.~\ref{fig:strange-charm-PDFs}.
Our imaginary matrix elements are proportional to the sum of the quark and antiquark distribution, and we clearly see that the strange contribution is about a factor of 5 or larger than the charm ones.
They are consistently smaller than those from CT18 and NNPDF3.1, possibly due to the missing contributions from the mixing with gluon matrix elements in the renormalization.
Higher statistics will be needed to better constrain the quark-antiquark asymmetry.
A full analysis of lattice-QCD systematics, such as finite-volume effects and discretization, is not yet included, and plans to extend the current calculations are underway.
The first nucleon gluon PDFs~\cite{Fan:2018dxu,Fan:2020cpa} and updated analysis can be found in this conference~\cite{Fan:2021Lat}.

\section{Meson PDFs}
\vspace{-0.3cm}
The first lattice-QCD calculation of the kaon valence-quark distribution functions using LaMET approach were reported in Ref.~\cite{Lin:2020ssv}.
The calculation is performed with multiple pion masses with the lightest one around 220~MeV, two lattice spacings $a=0.06$ and 0.12~fm, $(M_\pi)_\text{min} L \approx 5.5$, and high statistics ranging from 11,600 to 61,312 measurements.
We perform chiral-continuum extrapolation to obtain the renormalized matrix elements at physical pion mass, using a simple ansatz to combine our data from 220, 310 and 690~MeV: $h^R_{i}(P_z, z, M_\pi) = c_{0,i} + c_{1,i} M_\pi^2 + c_{a} a^2$ with $i=K, \pi$.
Mixed actions, with light and strange quark masses tuned to reproduce the lightest sea light and strange pseudoscalar meson masses, can suffer from additional systematics at $O(a^2)$;
such artifacts are accounted for by the $c_{a}$ coefficient.
We find all the $c_{a}$ to be consistent with zero.

\begin{figure}[htbp]
\includegraphics[width=.32\textwidth]{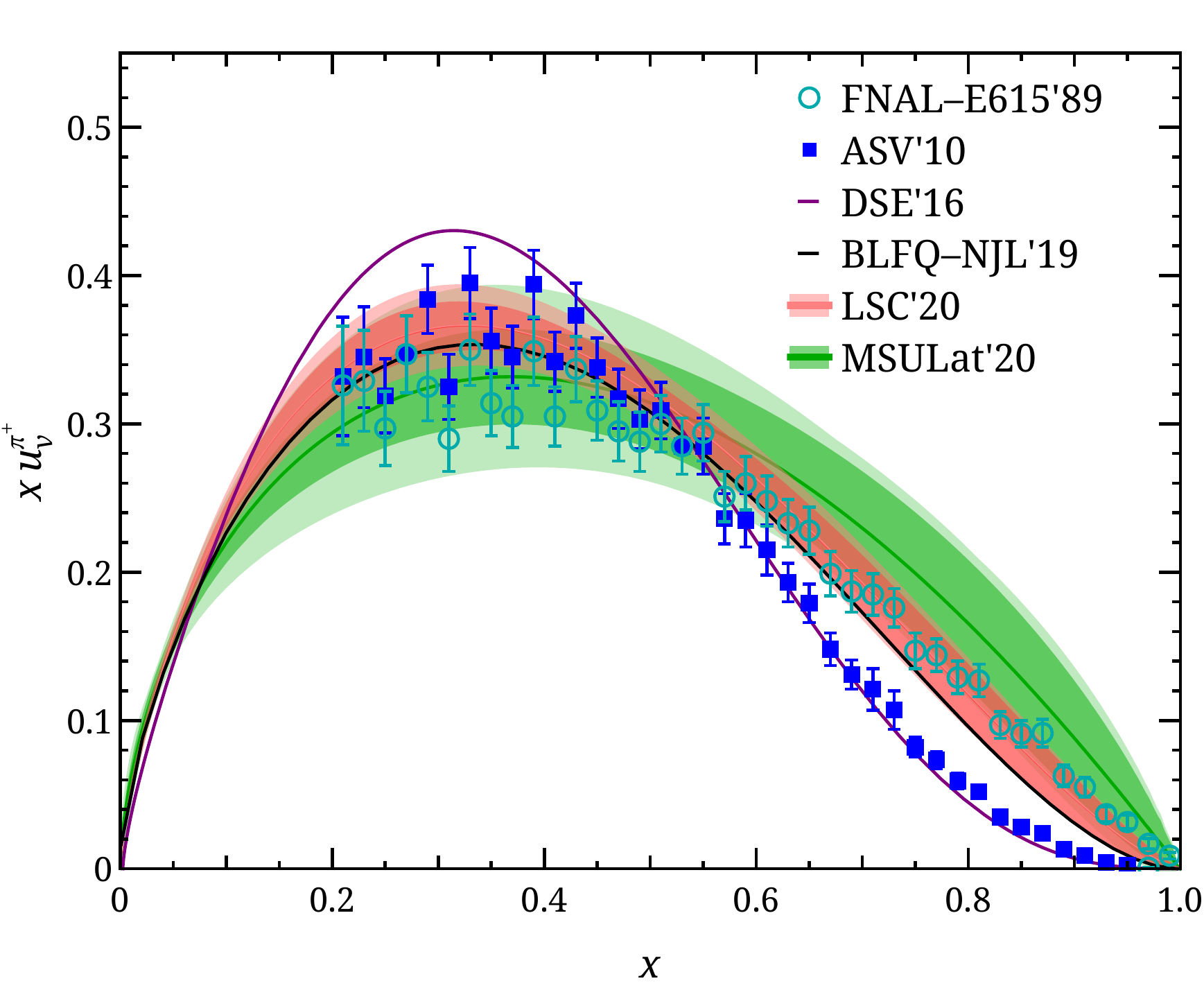}
\includegraphics[width=.33\textwidth]{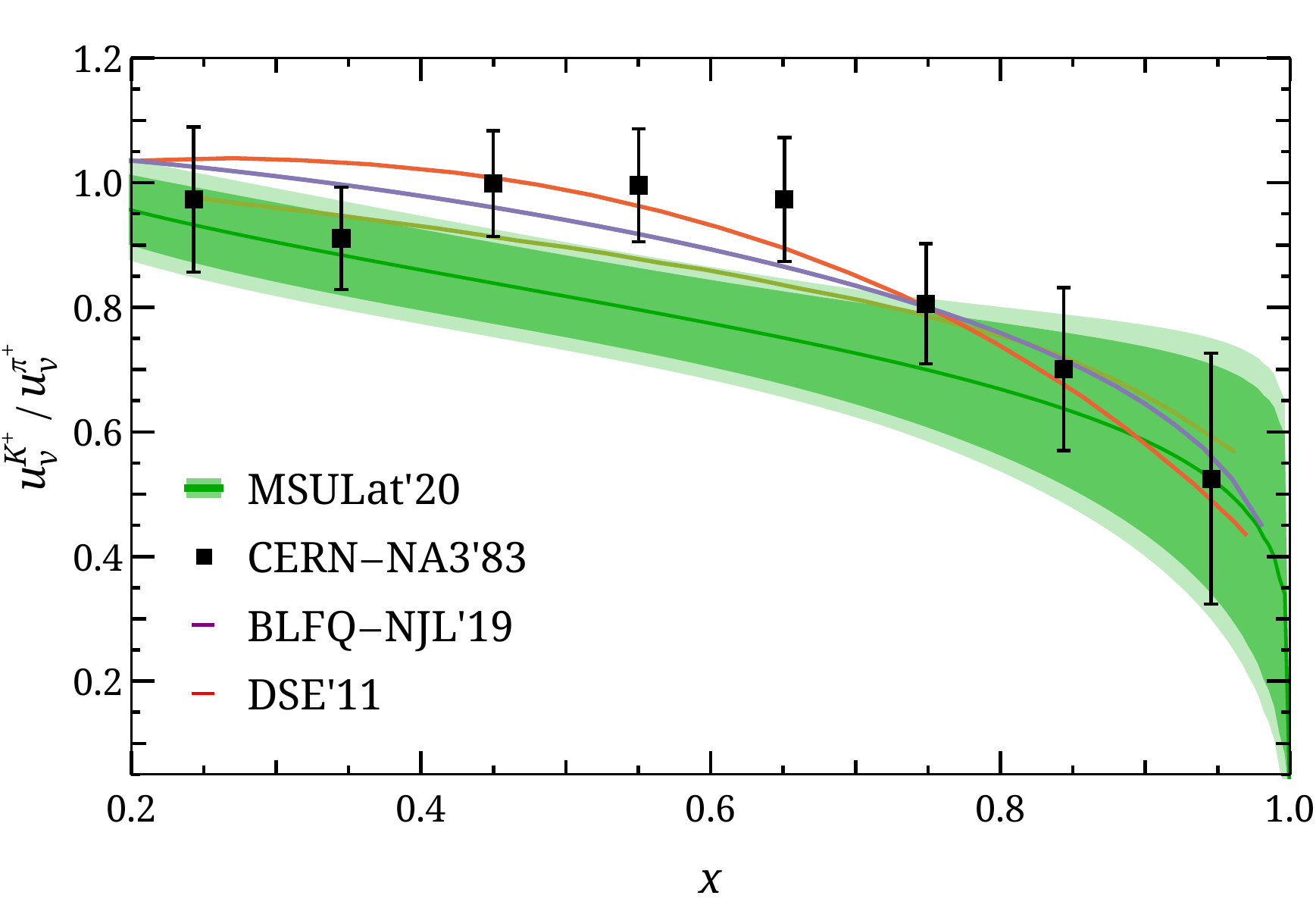}
\includegraphics[width=.32\textwidth]{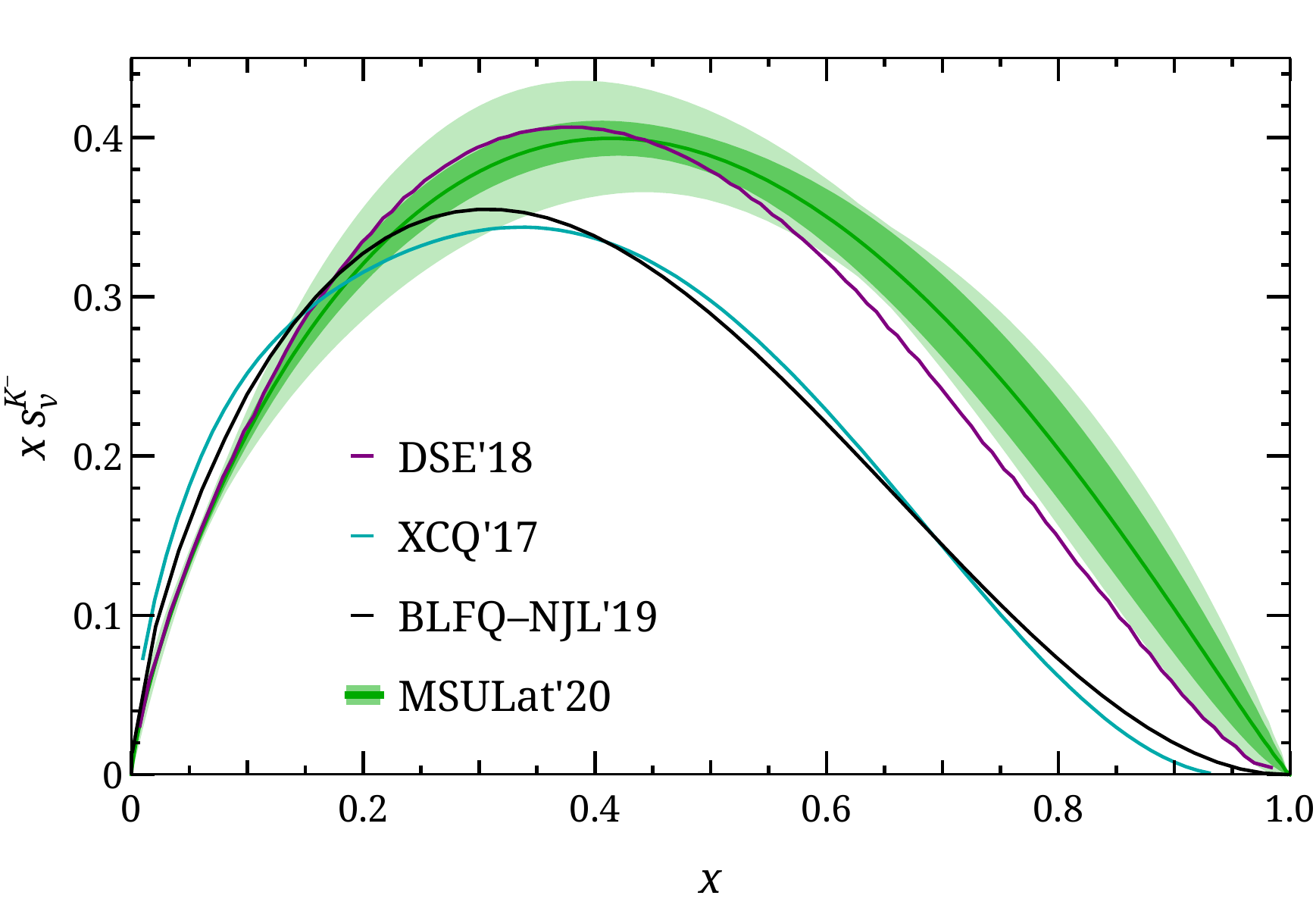}
\caption{Our results on the valence-quark distribution of the pion (left), the ratio of the light-quark valence distribution of kaon to that of pion (middle) and $x \overline{s}_v^K(x)$ as a function of $x$ (right) at a scale of $27\text{ GeV}^2$, both labeled ``MSULat'20'', along with results from relevant experiments and other calculations~\cite{Lin:2020ssv}.
The inner bands indicate statistical errors with the full range of $zP_z$ data while outter bands includes errors from using different data choices and fit forms.
}
\label{fig:mesonPDFs}
\end{figure}

Figure~\ref{fig:mesonPDFs} shows our final results for the pion valence distribution at physical pion mass ($u_v^{\pi^+}$) multiplied by Bjorken-$x$ as a function of $x$.
The inner bands indicate the statistical errors while the outer bands account for systematics errors from parametrization choices in the fits, the dependence on the maximum available Wilson-line displacement, etc.
We evolve our results to a scale of 27$\text{ GeV}^2$ using the NNLO DGLAP equations from the Higher-Order Perturbative Parton Evolution Toolkit (HOPPET) to compare with other results.
Our result approaches $x=1$ as $(1-x)^{1.01}$ and is consistent with the original analysis of the FNAL-E615 experiment data.
On the other hand, there is tension with the $x>0.6$ distribution from the re-analysis of the FNAL-E615 experiment data using next-to-leading-logarithmic threshold resummation effects in the calculation of the Drell-Yan cross section (labeled as ``ASV'10''), which agrees better with the distribution from Dyson-Schwinger equations (DSE)~\cite{Chen:2016sno};
both prefer the form $(1-x)^2$ as $x \to 1$.
An independent lattice study of the pion valence-quark distribution~\cite{Sufian:2020vzb}, also extrapolated to physical pion mass, using the ``lattice cross sections'' (LCSs), reported similar results to ours.

The middle of the Fig.~\ref{fig:mesonPDFs} shows the ratios of the light-quark distribution in the kaon to the one in the pion ($u_v^{K^+}/u_v^{\pi^+}$).
When comparing our result with the experimental determination of the valence quark distribution via the Drell-Yan process by NA3 Collaboration in 1982, we found good agreement between our results and the data.
Our result approaches $0.4$ as $x \to 1$ and agrees nicely with other analyses, such as constituent quark model,
the DSE approach (``DSE'11''),
and basis light-front quantization with color-singlet Nambu--Jona-Lasinio interactions (``BLFQ-NJL'19'').
Our prediction for $x s_v^{K}$ is also shown in Fig.~\ref{fig:mesonPDFs} with the lowest three moments of $s_v^{K}$ being $0.261(8)_\text{stat}(8)_\text{syst}$, $0.120(7)_\text{stat}(9)_\text{syst}$, $0.069(6)_\text{stat}(8)_\text{syst}$, respectively;
the moment results are within the ranges of the QCD-model estimates
from chiral constituent-quark model
(0.24, 0.096, 0.049)
and DSE~\cite{Chen:2016sno} (0.36, 0.17, 0.092).
The pion gluon PDF results can be found in Refs.~\cite{Fan:2021bcr,Fan:2021Lat}.

 \begin{table}
    \centering
    \begin{tabular}{|c|ccc|}
    \hline
$\langle x^n \rangle$ & $u_v^{\pi}$ & $u_v^{K^+}$ &  $s_v^{K^+}$\\
    \hline
$\langle x \rangle$ &  $0.225(18)_\text{stat}(10)_\text{syst}$
&  $0.192(8)_\text{stat}(6)_\text{syst}$  &  $0.261(8)_\text{stat}(8)_\text{syst}$ \\
     \hline
$\langle x^2 \rangle$ & $0.100(13)_\text{stat}(5)_\text{syst}$
&  $0.080(7)_\text{stat}(6)_\text{syst}$ & $0.120(7)_\text{stat}(9)_\text{syst}$ \\
    \hline
$\langle x^3 \rangle$ &  $0.056(10)_\text{stat}(2)_\text{syst}$%
& $0.041(6)_\text{stat}(4)_\text{syst}$ &  $0.069(6)_\text{stat}(8)_\text{syst}$ \\
    \hline
    \end{tabular}
\caption{Meson moments after continuum-physical extrapolation with statistical and systematic errors~\cite{Lin:2020ssv}.
}
    \label{tab:lat_info}
\end{table}

\section{Nucleon GPDs}
\vspace{-0.3cm}
The unpolarized (and polarized) GPDs $H(x,\xi,t)$ and $E(x,\xi,t)$ ($\tilde{H}(x,\xi,t)$ and $\tilde{E}(x,\xi,t)$) are defined in terms of the matrix elements
%
\begin{align}
F_q(x,\xi,t)&=\int\frac{dz^-}{4\pi}e^{ixp^+ z^-}\left\langle p''\left|\bar\psi\left(-\frac{z}{2}\right)\gamma^+ L\left(-\frac{z}{2},\frac{z}{2}\right)\psi\left(\frac{z}{2}\right)\right|p'\right\rangle_{z^+=0,\vec z_\perp=0}\nonumber\\
&=\frac{1}{2p^+}\left[H(x,\xi,t)\bar u(p'')\gamma^+ u(p')+E(x,\xi,t)\bar u(p'')\frac{i\sigma^{+\nu}\Delta_\nu}{2m}u(p')\right], \nonumber\\
\tilde F_q(x,\xi,t)&=\int\frac{dz^-}{4\pi}e^{ixp^+ z^-}\left\langle p''\left|\bar\psi\left(-\frac{z}{2}\right)\gamma^+ \gamma^5 L\left(-\frac{z}{2},\frac{z}{2}\right)\psi\left(\frac{z}{2}\right)\right|p'\right\rangle_{z^+=0,\vec z_\perp=0}\nonumber\\
&=\frac{1}{2p^+}\left[\tilde H(x,\xi,t)\bar u(p'')\gamma^+ \gamma^5 u(p')+\tilde E(x,\xi,t)\bar u(p'')\frac{\gamma^5\Delta^+}{2m}u(p')\right],
\end{align}
where $L(-z/2,z/2)$ is the gauge link along the lightcone and
$\Delta^\mu=p''^\mu-p'^\mu$
$t=\Delta^2$,
$\xi=\frac{p''^+-p'^+}{p''^+ +p'^+}$.
In the limit $\xi, t\to 0$, $H$ ($\tilde{H}$) reduces to the usual unpolarized (polarized) parton distributions.

\begin{figure}[tb]
\includegraphics[width=0.32\textwidth]{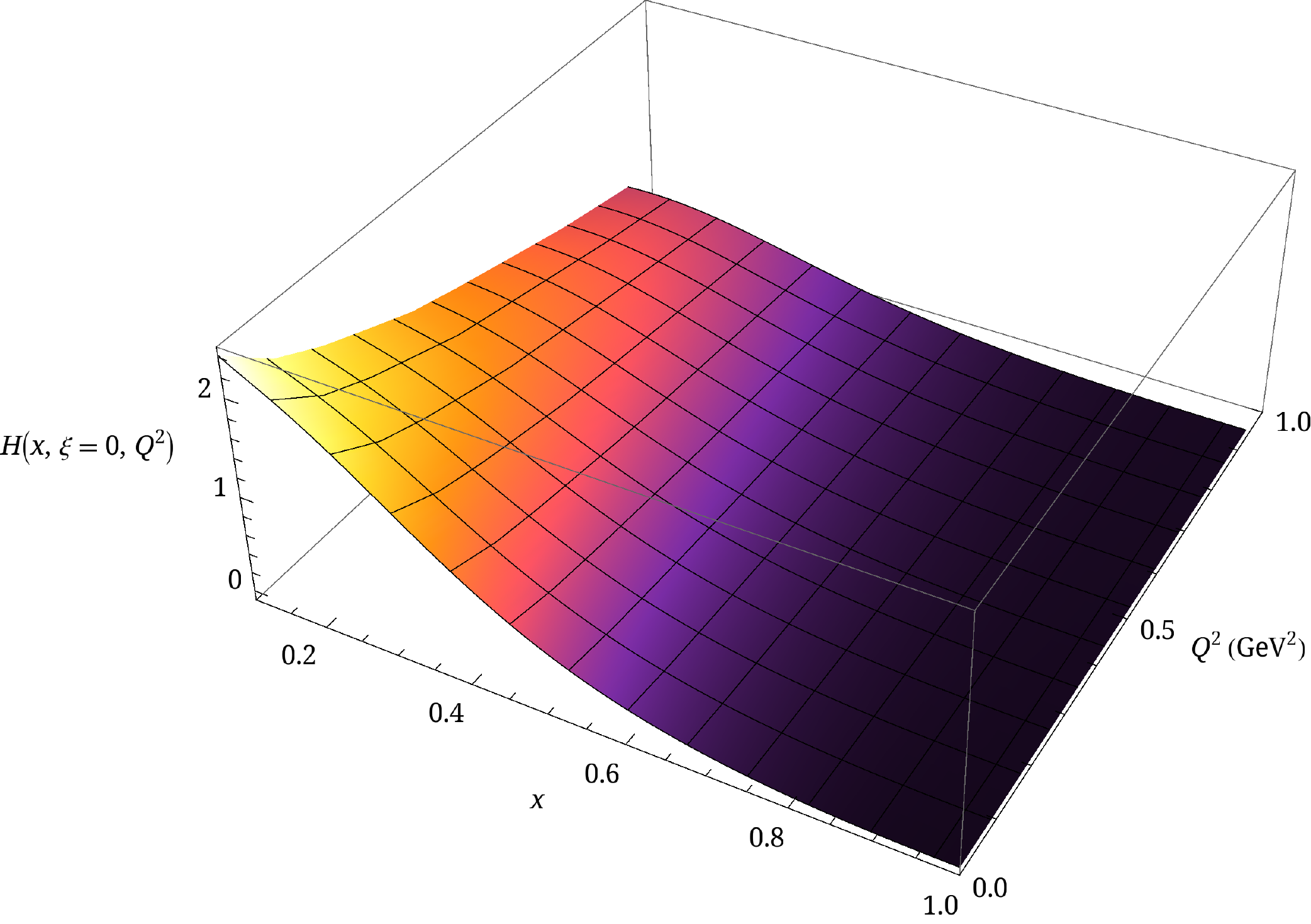}
\includegraphics[width=0.32\textwidth]{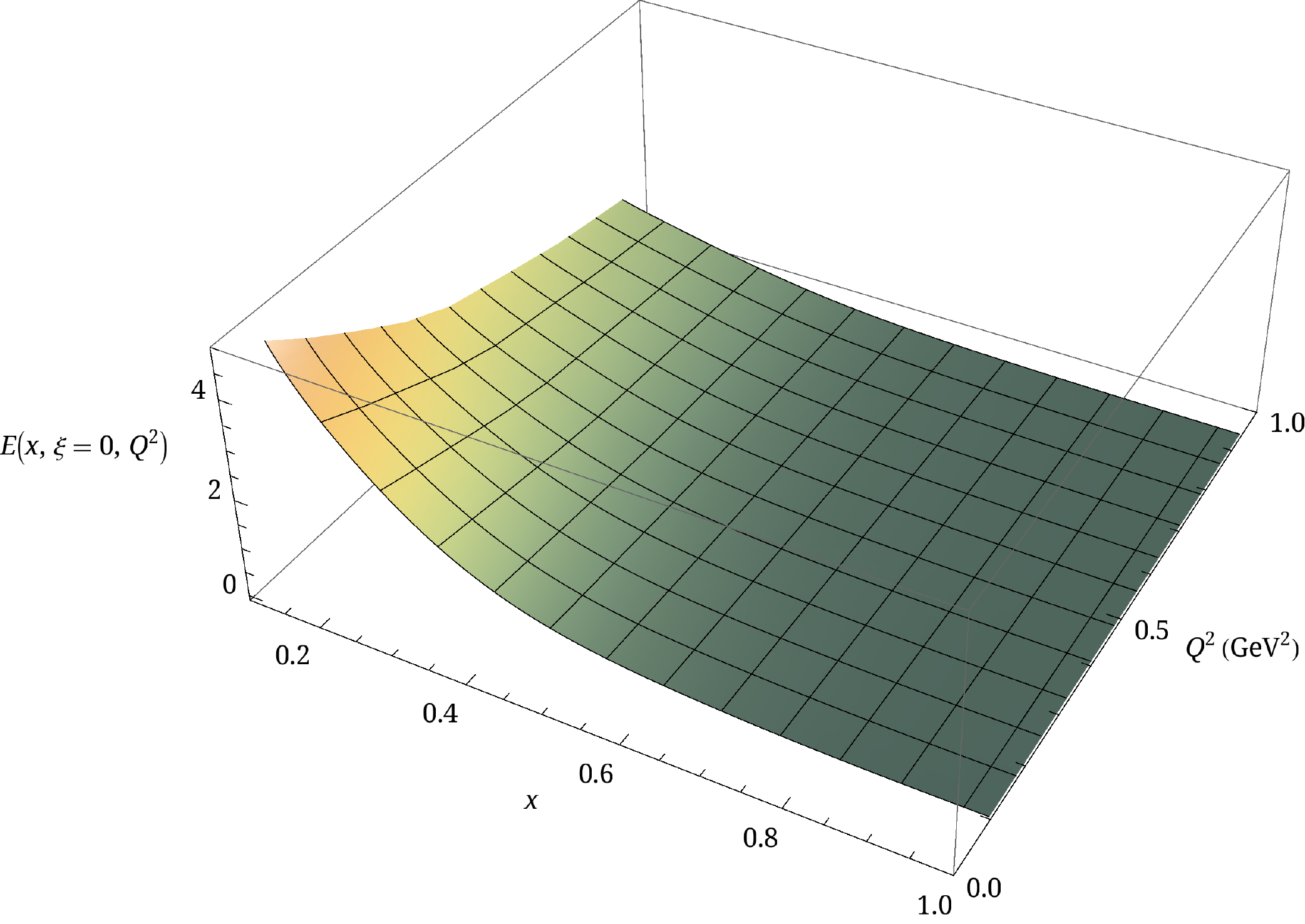}
\includegraphics[width=0.32\textwidth]{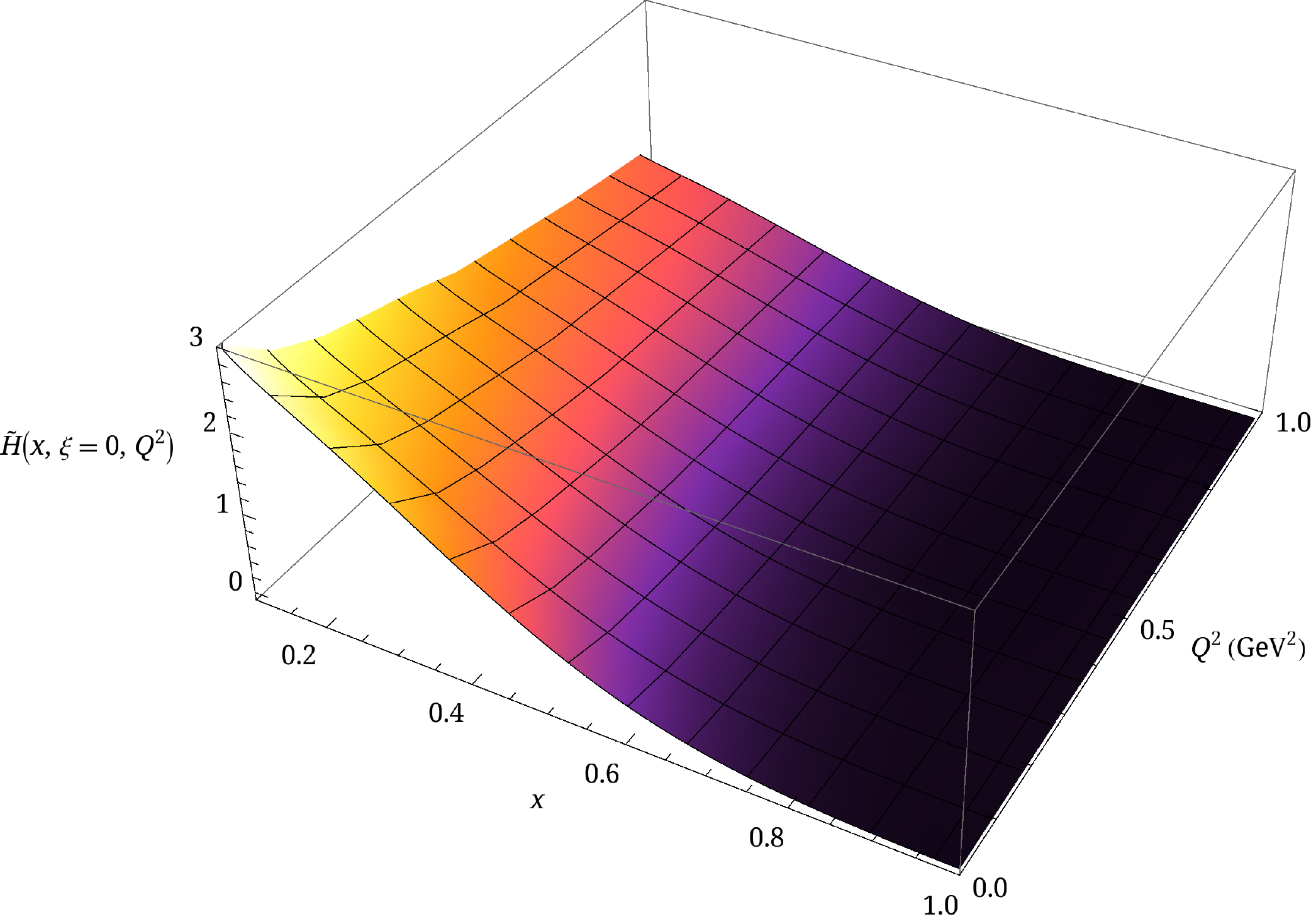}
\caption{
Nucleon isovector $H$ (left), $E$ (middle),  $\tilde{H}$ (right) GPDs at $\xi=0$ as functions of $x$ and momentum transfer $Q^2$.
\label{fig:3DGPD}}
\end{figure}

This is the first lattice calculation at physical pion mass of the nucleon isovector unpolarized GPDs.
It is performed using a lattice ensemble with 2+1+1 flavors of highly improved staggered quarks (HISQ) generated by MILC Collaboration, with lattice spacing $a\approx 0.09$~fm and volume $64^3\times 96$.
We use momentum-smeared sources to improve the signal at nucleon boost momentum $P_z \approx 2.2$~GeV, and report results at nonzero momentum transfers in $[0.2,1.0]\text{ GeV}^2$, focusing unpolarized GPDs and their quasi-GPD counterparts defined in terms of spacelike correlations calculated in Breit frame.
Nonperturbative renormalization in RI/MOM scheme is used to obtain the quasi-distribution before matching to the lightcone GPDs~\cite{Ji:2015qla,Liu:2019urm}.
The three-dimensional distributions $H(x,Q^2)$ and $E(x,Q^2)$ at $\xi=0$ are presented, along with the three-dimensional nucleon tomography and impact-parameter--dependent distribution for selected Bjorken $x$ at $\mu=3$~GeV in $\overline{\text{MS}}$ scheme,
as shown in Fig.~\ref{fig:3DGPD}.

\begin{figure}[tb]
\centering
\includegraphics[width=0.32\textwidth]{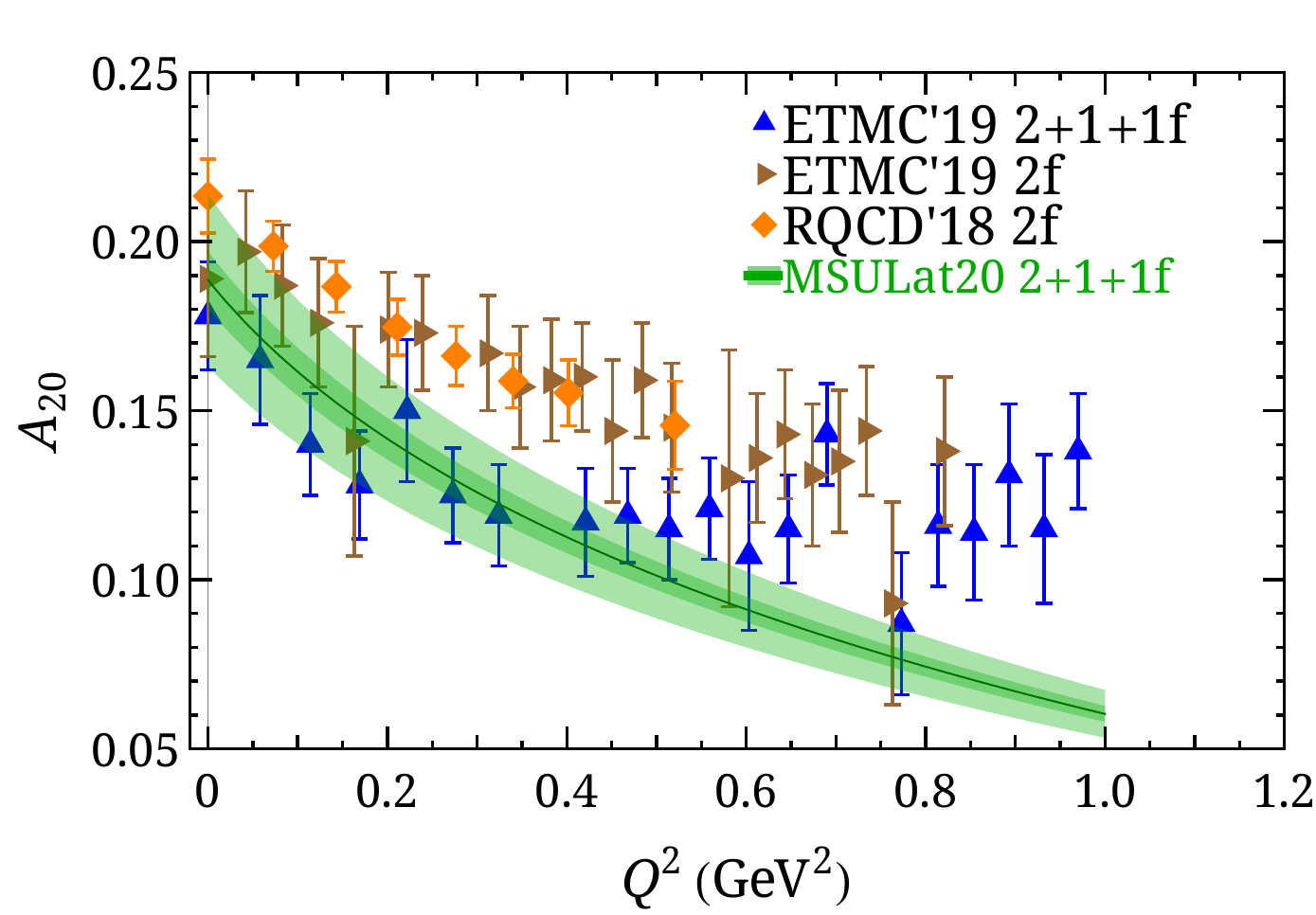}
\includegraphics[width=0.32\textwidth]{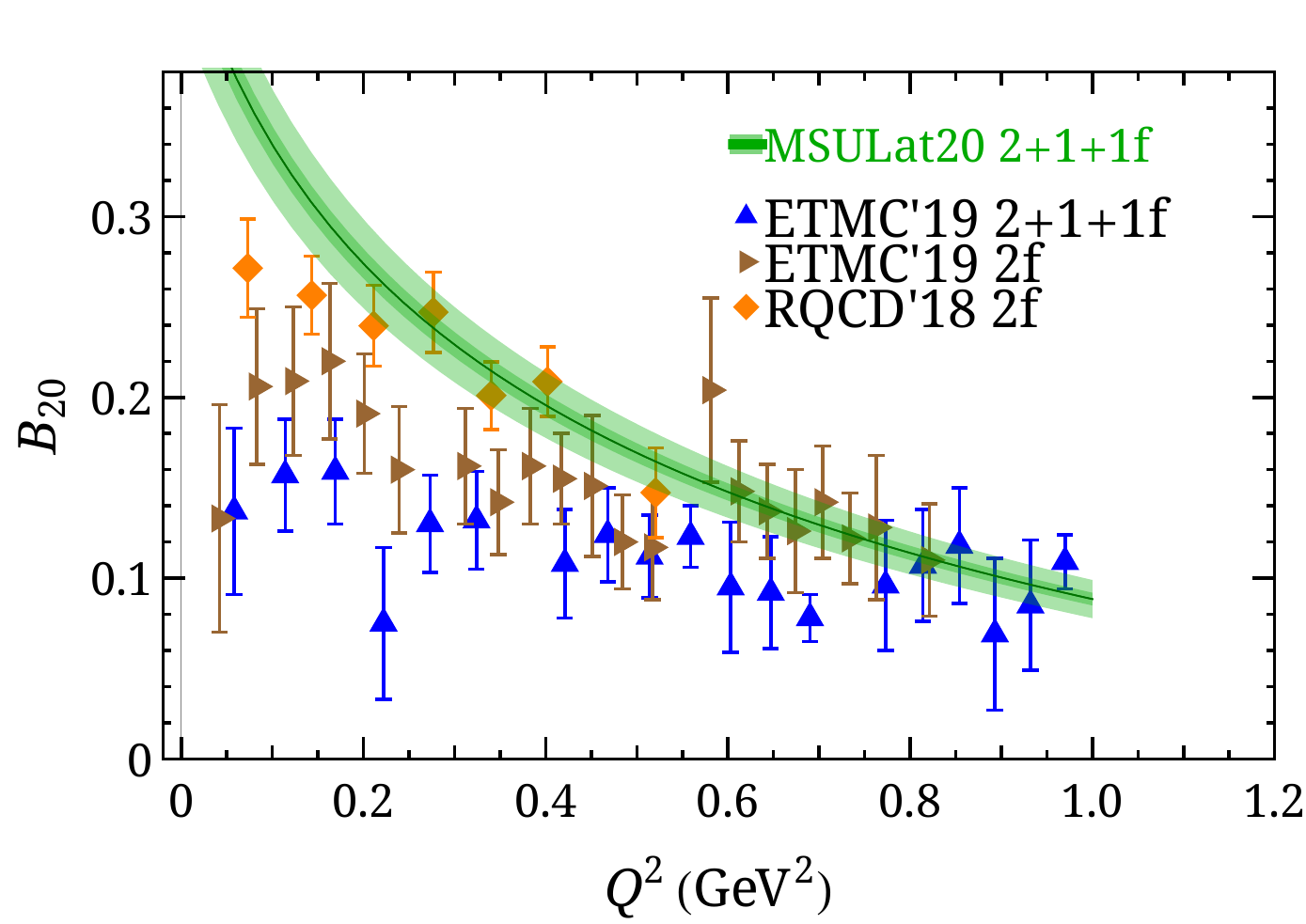}
\includegraphics[width=0.32\textwidth]{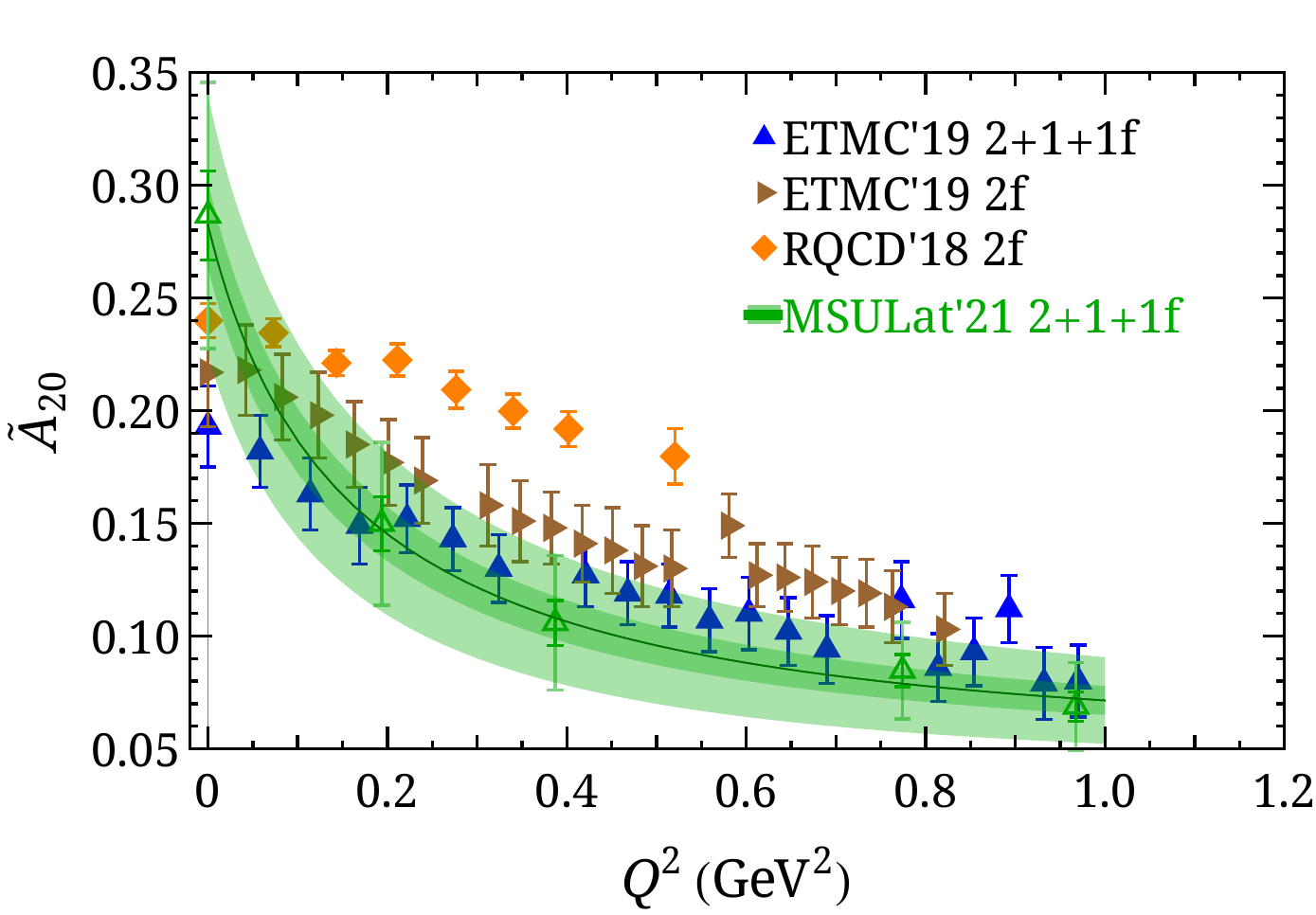}
\caption{
\label{fig:LatGFF}
The unpolarized nucleon isovector GFFs ($A_{20}$, $B_{20}$) and preliminary results on the linearly polarized nucleon isovector GFF $\tilde{A}_{20}(Q^2)$ obtained from MSULat using LaMET approach, compared with lattice OPE results calculated near physical pion mass as functions of transfer momentum $Q^2$:
ETMC19~\cite{Alexandrou:2019ali},
and RQCD19~\cite{Bali:2018zgl}.
}
\end{figure}

In the $\xi\to 0$ limit, the $H$ and $E$ GPDs decrease monotonically as $x$ or  $Q^2$ increases.
We take Mellin moments of the GPDs to compare with previous lattice calculations done using local matrix elements through the operator product expansion (OPE).
Taking the $x$-moments of $H$ and $E$:
\begin{subequations}
\begin{align}
\label{eq:GFFs}
\int_{-1}^{+1}\!\!dx \, x^{n-1} \, H(x, \xi, Q^2) =
\sum\limits_{i=0,\text{ even}}^{n-1} (-2\xi)^i A_{ni}(Q^2) &+ (-2\xi)^{n} \,  C_{n0}(Q^2)|_{n\text{ even}}, \\
\int_{-1}^{+1}\!\!dx \, x^{n-1} \, E(x, \xi, Q^2) =
\sum    \limits_{i=0,\text{ even}}^{n-1}	(-2\xi)^i B_{ni}(Q^2)	 &- (-2\xi)^{n} \,  C_{n0}(Q^2)|_{n\text{ even}}\,,
\end{align}
\end{subequations}
where the generalized form factors (GFFs) $A_{ni}(Q^2)$, $B_{ni}(Q^2)$ and $C_{ni}(Q^2)$ in the $\xi$-expansion on the right-hand side are real functions.
Our GFFs results using LaMET approach, with statistical (inner band) and systematic (outer band) are in reasonable agreements with lattice OPE method done at near physical pion mass ensembles, as shown in Fig.~\ref{fig:LatGFF}.
This demonstrates that LaMET methods are well suited for future GPD calculations.

\providecommand{\href}[2]{#2}\begingroup\raggedright\endgroup

\end{document}